\documentclass[12pt]{article} 
\usepackage{graphicx,subfigure,latexsym,amssymb}
\usepackage{bm}
\usepackage{amsmath}

\newcommand{\be}{\begin{equation}}
\newcommand{\ee}{\end{equation}}
\newcommand{\bear}{\begin{eqnarray}}
\newcommand{\eear}{\end{eqnarray}}
\newcommand{\ba}{\begin{array}}
\newcommand{\ea}{\end{array}}

\def\({\left(}
\def\){\right)}

\begin{document}

\begin{titlepage}

\vfill
\begin{center}
{\Large\bf Quantum Kinetic Theory of Spin Polarization of Massive Quarks in Perturbative QCD: Leading Log }

\vskip 0.3in
Shiyong Li\footnote{e-mail: {\tt  sli72@uic.edu}} and
Ho-Ung Yee\footnote{e-mail:
{\tt hyee@uic.edu}}
\vskip 0.3in

 {\it Department of Physics, University of Illinois,} \\
{\it Chicago, Illinois 60607 }
\\[0.3in]

{\normalsize  2019}

\end{center}

\begin{abstract}

We present the quantum kinetic equation for spin polarization of massive quarks in leading log order of perturbative QCD, which describes time evolution of the spin density matrix in momentum space of a massive quark interacting with a background QCD plasma. We find that the time evolution operator of the spin density matrix, or the quantum kinetic collision terms, are universally of order $\alpha_s^2\log(1/\alpha_s)$ in terms of the QCD coupling constant $\alpha_s=g^2/(4\pi)$.  Our quantum kinetic equation is valid for an arbitrary quark mass $m\gg m_D\sim gT$, where $m_D$ is the Debye mass, and can be used to study relaxation dynamics of spin polarization of massive quarks in perturbative QCD regime.

\end{abstract}

\vfill

\end{titlepage}
\setcounter{footnote}{0}

\baselineskip 18pt \pagebreak
\renewcommand{\thepage}{\arabic{page}}
\pagebreak

\section{Introduction \label{sec1}}

The quark-gluon plasma created in off-central heavy-ion collisions is expected to have a sizable collective fluid rotation that is 
originated from the large initial angular momentum of the two colliding projectiles. A part of the orbital angular momentum that resides 
in this fluid motion, or vorticity, will then be transferred to spin angular momenta of quasi-particles by interactions \cite{Liang:2004ph,Gao:2007bc,Betz:2007kg,Becattini:2007sr,Huang:2011ru,Becattini:2013vja,Pang:2016igs,Jiang:2016woz,Sun:2017xhx,Shi:2017wpk,Wei:2018zfb,Xia:2018tes}.
In equilibrium, the resulting spin-dependent distribution function of quasi-particles can be shown to be equal to what 
one would have as if the energy of a particle of spin $\bm S$ was shifted by $\Delta E=-\bm S\cdot \bm\omega$ in thermal distribution\cite{Becattini:2007zn,Becattini:2013fla,Florkowski:2018ahw}, where $\bm\omega={1\over 2}\bm\nabla\times\bm v$ is the vorticity of fluid\footnote{An intuitive derivation of this fact, based on a detailed balance argument with total angular momentum conservation was given in Ref.\cite{Chen:2014cla}.}.
Since $\bm S\sim {\cal O}(\hbar)$, the net spin angular momentum is quantum in nature, and is parametrically small (by $\hbar$) compared to macroscopic orbital angular momentum of the fluid.
The recent experimental observation of spin polarization of $\Lambda$ baryons in off-central heavy-ion collisions \cite{STAR:2017ckg} confirms the existence of this phenomenon that involves quantum spin of quasi-particles.

 In the early stage of heavy-ion collision, the QCD plasma is presumably in its de-confined phase, where quarks and gluons are the basic degrees of freedom.  The magnitudes of vorticity and magnetic field that polarize the quasi-particle spin are strongest at such an early stage.  Some of the spin polarization of quarks and gluons in this phase can be transferred to that of hadrons after hadronization, which may affect the experimentally observed spin polarization of hadrons. Whether this effect survives hadronic phase dynamics depends on the relaxation dynamics of spin polarization in hadronic phase, as well as many other realistic conditions of heavy-ion collisions \cite{Xia:2019fjf,Becattini:2019ntv}. As a first step, a reliable tracking of time-evolution of spin polarization of quarks and gluons within the de-confined phase itself would be a prerequisite in any quantitative theory prediction of spin polarization of observed hadrons. The aim of the present work is to address this problem, at least partly, in leading log order of perturbative QCD (pQCD). This also complements our previous work on the similar question 
in strongly coupled regime described by AdS/CFT correspondence \cite{Li:2018srq}.

In a time-dependent background such as heavy-ion collisions, the spin polarization of quasi-particles would naturally be driven off equilibrium. The time evolution of spin polarization would roughly be a competition between QCD dynamics that tries to relax the spin to equilibrium and the time-variation of backgrounds, such as vorticity and magnetic field, that drives the spin polarization of the system off equilibrium. 
If the time-variation of background is much slower than the characteristic relaxation time due to QCD interactions (which turns out to be $\tau_R\sim (\alpha_s^2\log(1/\alpha_s) T)^{-1}$), the system would follow closely the instantaneous equilibrium at each time. On the opposite case, the system would deviate significantly from equilibrium, and the spin polarization should be determined by solving the dynamical equation for time-evolution of spin polarization.

The present work is a small step in formulating such a dynamical equation of spin polarization of quasi-particles in QGP phase in leading log order of pQCD, focusing only on the spin polarization of massive quark, that may be suitable for strange quark or more massive quark species. 
Specifically, we assume that the mass is of hard-scale, $m\gg m_D\sim gT$ ($g$ is the QCD coupling constant). We will see that this justifies a few simplifications we will detail below. 
Therefore, our results would not be applicable for light (u,d) quarks and gluons, the study of which
we defer to a future work.

Schematically, the evolution of spin density matrix $\hat\rho$ of massive quark would take a form in linear order as (basically a ``Lindblad equation")
\be
{\partial \hat\rho\over\partial t}=-\Gamma\cdot\hat\rho  -{i\over\hbar}[H_{\rm eff},\hat\rho]\,,\label{eqS}
\ee
with a linear relaxation operator $\Gamma$, and the effective one-particle Hamiltonian $H_{\rm eff}$ in a 2-dimensional spin space, that may include vorticity, $\bm\omega$, and magnetic field, $\bm B$, in somewhat phenomenological way as
\be
H_{\rm eff}=-{\hbar\over 2}\bm\sigma\cdot (\bm\omega+Q\bm B)\,,
\ee
where $Q$ is the electromagnetic charge of quark.
In general, the $\Gamma$ should also depend on vorticity and magnetic field in such a way that the equilibrium spin density matrix is given by $\hat\rho_{eq}=e^{-H_{\rm eff}/T}\approx {\bf 1}-H_{\rm eff}/T$, at least in linear order in $\bm \omega$ and $\bm B$. The magnetic field, for example, should modify the wave function and energy spectrum of quark states in the computation of $\Gamma$ that we describe in the following sections. We would expect to have an expansion of $\Gamma$ in small $\bm\omega$ and $\bm B$ as $\Gamma=\Gamma_0+\Gamma_1+\cdots$ where $\Gamma_1$ is linear in $\bm\omega$ or $\bm B$.
In this work, we present our result for the leading relaxation operator $\Gamma_0$, that corresponds to the case of vanishing $\bm\omega$ and $\bm B$, and the computation of $\Gamma_1$ is planned in the future.

Although $\Gamma_0$ is not sufficient to describe the spin polarization in a time-varying vorticity and magnetic field, it can still be used to compute spin-related correlation functions in a linear response theory. At least, $\Gamma_0$ can describe how a spin density matrix, initially polarized, relaxes to the unpolarized one (the identity operator in spin space), when vorticity and magnetic field cease to exist.

In general, the density matrix $\hat\rho$ is defined in the phase space $(\bm x,\bm p)$ in addition to spin space.
A convenient way to think about it is in the language of Schwinger-Keldysh contour. The position and momentum operators in forward and backward time contours (labeled as 1 and 2 respectively) satisfy the commutation relations $[\bm x^i_1,\bm p^j_1]=i\hbar\delta^{ij}$ and $[\bm x^i_2,\bm p^j_2]=-i\hbar\delta^{ij}$. In terms of ``ra'' variables where $r={1\over 2}(1+2)$ and $a=1-2$,
the only non-vanishing commutators are $[\bm x^i_{r/a},\bm p^j_{a/r}]=i\hbar\delta^{ij}$, especially $\bm x_r$ and $\bm p_r$ commute with each other. This allows us to introduce a wave function $\hat\rho(\bm x_r,\bm p_r)$ which is our density matrix in phase space. Since $x_r$ and $p_a$ are conjugate variables, the wave function (or density matrix) in momentum space $\hat\rho(\bm p_r,\bm p_a)=\hat\rho(\bm p_1,\bm p_2)$ is related to the density matrix in phase space $\hat\rho(\bm x_r,\bm p_r)$ by a Fourier (or Wigner) transform 
\be
\hat\rho(\bm x_r,\bm p_r)=\int {d^3\bm p_a\over (2\pi)^3}\,e^{i\bm p_a\cdot\bm x_r}\,\hat\rho(\bm p_r,\bm p_a)\,.\label{wigner}
\ee
We will assume that the density matrix in phase space $\hat\rho(\bm x,\bm p)$ is a slowly varying function on space $\bm x$,
compared to a microscopic scale of QCD interactions, usually set by the mean free path $l_{mfp}\sim (\alpha_s^2 T)^{-1}$.
This means that we can consider $\bm x_r$ as constant in the computation of relaxation operator $\Gamma$ in (\ref{eqS}).
This is translated to a smallness of $\bm p_a\sim \partial_{\bm x}\ll l_{mfp}^{-1}$ by (\ref{wigner}), that is, the density matrix in momentum space $(\bm p_1,\bm p_2)$ is nearly diagonal in momentum variables. If we neglected spin degrees of freedom, these diagonal elements would correspond to a usual distribution function $f(\bm p)$ in momentum space.
 In our computation of $\Gamma$, we therefore work with diagonal elements in the density matrix in momentum space, defined by $\hat\rho(\bm p_1,\bm p_2)\equiv (2\pi)^3\delta(\bm p_1-\bm p_2)\hat\rho(\bm p_1)$. This is justified as long as we don't care about the advective terms in $\bm x$ in quantum kinetic equation, but focus only on local ``collision terms'' of $\Gamma$ in (\ref{eqS}) in the {\it spatial homogeneous limit}.
Note that we still keep a full spin matrix of $\hat\rho(\bm p)$ in the spin space.

The reason why we need to keep full quantum correlation of spin degrees of freedom in the density matrix $\hat\rho(\bm p)$
is that the two spin states are degenerate in energy and the quantum correlation time is arbitrarily large $\tau_q\sim {\hbar\over\Delta E}\to\infty$. Even in the presence of background vorticity or magnetic field, the energy shift is $\Delta E\sim \bm S\cdot\bm\omega$ or $\Delta E\sim Q\bm S\cdot\bm B$ ($Q$ is charge) which is ${\cal O}(\hbar)$ since $\bm S\sim {\cal O}(\hbar)$.
The quantum correlation time for spin is $\tau_q\sim {\hbar\over \Delta E}\sim {\cal O}(\hbar^0)$, and  it is in the classical time scale, that is usually described by a classical Boltzmann kinetic theory. Therefore, quantum correlation of spin should be considered even in a regime of kinetic theory, and hence ``quantum kinetic theory".

A more fundamental treatment of (\ref{eqS}), that should require a significantly larger effort in the future, would involve a complete analysis of spin density matrix in the full phase space $(\bm x,\bm p)$. The free streaming, collision-less quantum kinetic equation for this case was recently studied in Ref \cite{Mueller:2019gjj,Weickgenannt:2019dks,Gao:2019znl,Hattori:2019ahi}. Our study in this sense can be viewed as providing the collision term in leading log of pQCD. However, our result should be improved in this case, including spatial gradient effects in the collision terms (we are restricting to the homogeneous limit, as described in the previous paragraph).
This is because the vorticity is a spatial gradient of background fluids.
Another way to understand this is that the orbital angular momentum of background fluids can in general be transferred to spin angular momentum of the massive quarks we are looking at. 
In this more fundamental picture, since total angular momentum has to be conserved, a loss of spin angular momentum of the massive quark (described by $\Gamma$) must be compensated by a gain of angular momentum in the background fluid. This gain term will be shared among all quasi-particles of the background, dominantly light quarks and gluons. 
Compared to the massive quark we consider, these other degrees of freedom is much larger, and the spin gain is diluted and its back reaction to the equation for $\hat\rho$ will be suppressed compared to the loss term.
More importantly, the gain in angular momentum of background will be shared between orbital and spin angular momenta.
Since spin is smaller than orbital by $\hbar$, the most of gain will go to the orbital angular momentum with a change of vorticity, $\Delta\bm\omega$. Since the spin is $\bm S\sim {\cal O}(\hbar)$, the change $\Delta\bm\omega\sim {\cal O}(\hbar)$, and its effect to the dynamics of $\bm S$ via $\Delta E=-\bm S\cdot\bm\omega\sim {\cal O}(\hbar^2)$ is higher order in small $\hbar$.
Based on this consideration, we neglect possible ``gain terms" in our quantum kinetic equation (\ref{eqS}). In essence, we treat the background as a spin reservoir that can absorb any change of spin angular momentum in $\hat\rho$ of ``dilute" massive quarks, without any back reaction of the absorbed spin to the evolution of $\hat\rho$ itself.
This is justified as long as we track the spin polarization of dilute massive quarks only, without caring about those of light quarks and gluons.

\begin{figure}[t]
 \centering
 \includegraphics[height=5.5cm]{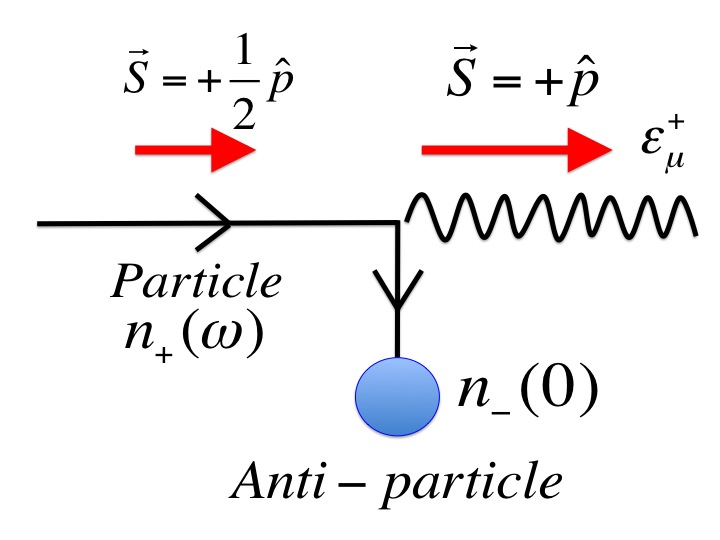}\caption{The quark-gluon conversion process that we can neglect for massive quarks in leading log order. \label{fig1}}
 \end{figure}

The characteristic relaxation rate of spin polarization of massive quarks will be shown to be of order $\Gamma_0 \sim \alpha_s^2\log(1/\alpha_s)T$ where $\alpha_s=g^2/(4\pi)$\footnote{It can be shown that this is in fact true for light quarks and gluons as well, as it is universal for all soft $t$-channel processes \cite{Arnold:2000dr}.}. This is of the same microscopic relaxation rate that governs other transport coefficients, such as shear viscosity or charge conductivities. See Refs.\cite{Grabowska:2014efa,Manuel:2015zpa} for similar observations, but in terms of usual scattering rate picture, i.e. considering only diagonal elements of the density matrix. These contributions to $\Gamma_0$ arise from soft $t$-channel gluon exchange of momentum $q$ in the scatterings with background hard thermal particles, where the log comes from a range $m_D\sim gT\ll q\ll T$. If the quark was light, there would also exist soft $t$-channel quark exchange contribution of the same leading log order, making conversion of a quark to a gluon \cite{Arnold:2000dr,Hong:2010at}. See Figure \ref{fig1}. 
The spin polarization of light quarks can be transferred to that of gluons and vice versa by these conversion processes, and a complete picture for light quarks would have to involve the spin density matrices of both light quarks and gluons.
For the process of Figure \ref{fig1} to happen, the exchanged quark should be the same species
of the incoming quark.
Our assumption of a hard-scale mass $m\gg gT$ for the massive quark implies $q\gtrsim m\gg gT$, which makes a soft $q$ quark exchange impossible for the massive quark. This justifies the absence of quark-gluon conversion process of Figure \ref{fig1} for massive quark at leading log in our study.

\section{ Time evolution of spin density matrix in Schwinger-Keldysh formalism \label{sec2}}

We consider the Hilbert space of one-quark state interacting with background QCD plasma degrees of freedom in a finite temperature $T$. This reduced description is justified as long as the occupation number of quark per unit quantum state (given by $({\rm Number\, of\, quarks})\times(2\pi\hbar)^3/(d^3\bm x d^3\bm p)$ in phase space, usually called the distribution function $f(\bm x,\bm p)$) is much less than unity (or ``dilute'' Boltzmann limit),
so that quantum statistics of Pauli blocking is negligible. We assume that our massive quark species satisfies this condition, either by $m\gtrsim T$ due to thermal Boltzmann suppression, or at least in early stages in heavy-ion collisions when the massive quarks are scarce. 
A convenient basis of states for our purpose is $\{|\bm p,\pm\rangle\}$ of a momentum $\bm p$ and helicity $h=\pm 1/2$ (meaning that the spin state is an eigenstate of the spin angular momentum
along $\hat{\bm p}=\bm p/p$ ($p\equiv |\bm p|$) with the eigenvalue $\pm \hbar/2$, that is, $(\hat{\bm p}\cdot \bm\sigma )|\bm p,\pm\rangle=\pm |\bm p,\pm\rangle$). 
As explained in the introduction, we consider a density matrix that is (approximately) diagonal in the momentum variable, that is sufficient for describing the local quantum collision term $\Gamma_0$
in the quantum kinetic equation. We therefore have the density matrix per unit volume as (we set $\hbar=1$ from here without much confusion)
\be
\hat\rho=\int {d^3\bm p\over (2\pi)^3 } \hat\rho(\bm p)\,,\label{densityinp}
\ee
where $\hat\rho(\bm p)$ is a $2\times 2$ spin density matrix at a fixed momentum $\bm p$.
More explicitly, we have
\be
\hat\rho(\bm p)=\sum_{s,s'=\pm} |\bm p,s\rangle \rho_{s,s'}(\bm p)  \langle \bm p,s'|\,,\label{helicitybasis}
\ee
in bra-ket notation, with a set of four functions in momentum space, $\rho_{s,s'}(\bm p)$.

It is important to recall a phase ambiguity of the basis states $|\bm p,s\rangle \to e^{i\phi(\bm p,s)}|\bm p,s\rangle$ with an arbitrary choice of $\phi(\bm p,s)$, which is reflected to the compensating phase ambiguity of $\rho_{s,s'}(\bm p)\to e^{-i(\phi(\bm p,s)-\phi(\bm p,s'))} \rho_{s,s'}(\bm p)$, such that the density matrix $\hat\rho$ is unambiguous. We will be careful about this ambiguity in our computation, such that our final quantum kinetic equation in terms of physical spin polarization is well defined free of this phase ambiguity.

One way of fixing the phase ambiguity is to work universally in the basis of z-component of spin operator.
Then, the helicity $s/2$ state has an explicit 2-component spinor representation, $\xi_s(\bm p)$, satisfying $(\hat{\bm p}\cdot\bm\sigma) \xi_s(\bm p)=s\xi_s(\bm p)$, with normalization $\xi_s^\dagger\xi_s=1$.
The density matrix in this basis is then an explicit $2\times 2$ matrix, given by
\be
\hat\rho(\bm p)=\sum_{s,s'=\pm} \xi_s(\bm p) \rho_{s,s'}(\bm p) \xi_{s'}^\dagger(\bm p)\,.\label{densitym}
\ee
We emphasize again that $\xi_s(\bm p)$ and hence $\rho_{s,s'}(\bm p)$ are each phase ambiguous, but the net density matrix $\hat\rho(\bm p)$ is free of ambiguity.
The spin operator in this basis is $\bm S={1\over 2}\bm\sigma$ (with $\hbar=1$), and the spin polarization density in momentum space from $\hat\rho(\bm p)$ is then given by
\be
\bm S(\bm p)={\rm Tr}(\bm S \hat\rho(\bm p))={1\over 2}{\rm Tr}(\bm \sigma \hat\rho(\bm p))\,.
\ee
Recalling that ${\rm Tr}(\hat\rho(\bm p))$ is the usual number distribution $f(\bm p)$ that appears in the conventional semi-classical Boltzmann equation, we can express the density matrix in this basis as
\be
\hat\rho(\bm p)={1\over 2}f(\bm p)+\bm S(\bm p)\cdot\bm\sigma\,.\label{phys}
\ee
Note that $f(\bm p)$ and $\bm S(\bm p)$ are physical quantities and are independent of our basis choice: we will present our quantum kinetic equation for $f(\bm p,t)$ and $\bm S(\bm p,t)$ ($t$ is time).
The total spin polarization and the number of quarks per unit volume is then given by
\be
\bm S=\int {d^3\bm p\over (2\pi)^3}\, \bm S(\bm p)\,,\quad n=\int {d^3\bm p\over (2\pi)^3}\, f(\bm p)\,.
\ee
 
The density matrix $\hat\rho(t)$ and its time evolution is most naturally described in the Schwinger-Keldysh formalism. The path integral in the time-forward contour (labeled as contour 1) gives the unitary time evolution of the ket part of the density matrix, and that in the time-backward contour (contour 2) gives the complex conjugate evolution of the bra part of the density matrix,
\be
\hat\rho(t)=\langle U_1(t,t_0)\hat\rho(t_0) U^\dagger_2(t,t_0)\rangle_A\,,\label{timeevol}
\ee
where $U_{1,2}(t,t_0)={\cal P}e^{-i\int_{t_0}^t dt' \, H_{1,2}(t')}$ are the unitary
time evolutions in the contours 1 and 2 respectively, and $\langle\cdots\rangle_A$ means the thermal path integral average of background degrees of freedom of the QCD plasma.
Since our system is an open system and is interacting with background degrees of freedom (more precisely, through the soft-scale color gauge field $A_\mu(x)$ in leading log order),
the $H_1(t)$ depends on the operator of the background degrees of freedom in contour 1 that couples our system to the background, and it is time-dependent in general due to time-dependence of that operator (i.e. the color gauge field in contour 1, $A^{(1)}_\mu(\bm x, t)$). The same is true for $H_2(t)$ and $A^{(2)}_\mu(\bm x,t)$. The average $\langle\cdots\rangle_A$ in the above then involves the thermal correlation functions of $A^{(1)}$ and $A^{(2)}$ in the Schwinger-Keldysh contours (the two-point functions in our leading order computation). 
In the frequency space, these correlation functions satisfy the KMS relations.
More explicitly, defining 
\be
G^{(ij)}_{\mu\nu}(q^0,\bm q)=\int d^3\bm x dt\,e^{i(q^0 t-\bm q\cdot\bm x)}\,\langle A^{(i)}_\mu(\bm x,t)A^{(j)}_\nu(\bm 0,0)\rangle_A \,,\quad i,j=1,2\,(\rm{SK\,contours})\label{corr}
\ee
what we will need later are the relations,
\be
G^{(12)}_{\mu\nu}(q^0,\bm q)=n_B(q^0) \rho_{\mu\nu}(q^0,\bm q)\,,\quad G^{(21)}_{\mu\nu}(q^0,\bm q)=(n_B(q^0)+1)\rho_{\mu\nu}(q^0,\bm q)\,,\label{KMS}
\ee
in terms of the gluon spectral density $\rho_{\mu\nu}\equiv i(G^R_{\mu\nu}-(G^R_{\nu\mu})^*)=-2\,{\rm Im}[G^R_{\mu\nu}]$, where $G^R$ is the retarded two-point function and the last equality holds only for symmetric case that is true in parity (P)-even background that we assume\footnote{See Refs.\cite{Jimenez-Alba:2015bia,Mamo:2015xkw} for an introduction to a possible anti-symmetric part, that is called ``P-odd spectral density'', in the presence of background axial charge.}, and $n_B(q^0)=1/(e^{\beta q^0}-1)$ is the Bose-Einstein distribution.
In our leading log computation, these correlation functions include the well-known 1-loop Hard-Thermal-Loop (HTL) self-energy, the imaginary part of which gives the non-vanishing spectral density in soft t-channel space-like momenta, that represents 
the scatterings with background thermal particles by cutting the 1-loop, while the real part regulates the infrared 
divergence in these t-channel scatterings by (real-time) screening effects due to background thermal particles (see the section \ref{secLL} for a more detailed review on this). 

The Hamiltonian in our one-quark picture is a sum of the free kinetic energy, $H_0$, and the QCD interaction with background gluon fields, $H_I$. The interaction Hamiltonian arises from the field theory Hamiltonian
\be
H_I=g\int d^3\bm x\, \bar\psi(\bm x)\gamma^\mu t^a \psi(\bm x) A^a_\mu(\bm x)\,,
\ee
where $\psi(\bm x)$ is the quark field operator, and $A^a_\mu(\bm x)$ is the gluon field with color index $a$ ($t^a$ are the color generators). We choose our convention as
\be
\gamma^0=\left(\begin{array}{cc} 0 & i{\bf 1}_{2\times 2}\\ i{\bf 1}_{2\times 2} & 0\end{array}\right)\,,\quad  \gamma^i=\left(\begin{array}{cc} 0 & i\sigma^i\\ -i\sigma^i &0\end{array}\right)\,, i=1,2,3\,,\label{gammaM}
\ee 
and $\bar\psi\equiv -\psi^\dagger\gamma^0$. In this convention, the quark spinor of momentum $\bm p$ and helicity $h=\pm 1/2$, that shares the same phase ambiguity as the state $|\bm p,\pm\rangle$, is explicitly given by
\be
|\bm p,s\rangle\sim u(\bm p,s)=\left(\begin{array}{c} \sqrt{E_p-s p}\,\xi_s(\bm p) \\ \sqrt{E_p+sp}\,\xi_s(\bm p)\end{array}\right)\,,\label{spinor}
\ee
where $p=|\bm p|$, $E_p=\sqrt{p^2+m^2}$ and $(\hat{\bm p}\cdot\bm\sigma)\xi_s(\bm p)=s\xi_s(\bm p)$ ($s=\pm 1$). This explicit expression will be used in our computation of spin-dependent transition amplitudes.
A quick way to see why the (arbitrary) phase of $|\bm p,s\rangle$ is identical to that of $u(\bm p,s)$ is to note the field operator $\psi(\bm x)$ expanded as
\be
\psi(\bm x)\sim \sum_{\bm p,s} u(\bm p,s) a_{\bm p,s} e^{i\bm p\cdot\bm x}+ {\rm h.c}
\ee
where $a_{\bm p,s}$ is the annihilation operator of one-quark state. Noting that $|\bm p,s\rangle\sim a^\dagger_{\bm p,s}|0\rangle$, the phase ambiguity of $|\bm p,s\rangle$ (or equivalently, $a^\dagger_{\bm p,s}$) is precisely identical to the phase ambiguity of $u(\bm p,s)$, such that
the $\psi(\bm x)$ operator and its conjugate entering the interaction Hamiltonian $H_I$ are unambiguous. Ultimately, this phase ambiguity becomes that of the 2-component spinor $\xi_s(\bm p)$ in (\ref{spinor}). As we use $H_I$ in our computation with a consistent use of $\xi_s(\bm p)$ in both $H_I$ and the definition of density matrix (\ref{densitym}), our result for $f(\bm p)$ and $\bm S(\bm p)$ is free of this ambiguity.

Since we need to keep the normalization of one-quark state correctly when discussing the density matrix, it is most convenient to work in a finite volume $V$ with discrete spectrum of states and then take an infinite volume limit. The momentum space becomes discrete $\bm p_{\bm n}$ with
integer-valued label vector $\bm n$, and the infinite volume limit is 
\be
\sum_{\bm n}\to V\int {d^3\bm p\over (2\pi)^3}\,.
\ee
The fields are expanded as (including only the quark sector, neglecting anti-quarks)
\be
\psi(\bm x)={1\over\sqrt{V}}\sum_{\bm n,s}{1\over \sqrt{2E_{p_{\bm n}}}} u(\bm p_{\bm n},s) e^{i\bm p_{\bm n}\cdot\bm x} a_{\bm p_{\bm n},s}\,,
\ee
and 
\be
A_\mu(\bm x,t)={1\over V}\sum_{\bm n} A_\mu(\bm p_{\bm n},t) e^{i\bm p_{\bm n}\cdot\bm x}+{\rm h.c.}\,,
\ee
where $a_{\bm p_{\bm n},s}$ is the annihilation operator of one-quark state $|\bm p_{\bm n},s\rangle$, with {\it unit normalization}, that is 
\be
\{a_{\bm p_{\bm n},s}, a^\dagger_{\bm p_{\bm n'},s'}\}=\delta_{\bm n,\bm n'} \delta_{s,s'}\,,
\ee
and $A_\mu(\bm p_{\bm n},t)$ is defined such that it has the two-point correlation functions as
\be
\langle A^{(i)}_\mu(\bm p_{\bm n},t) A^{(j)}_\nu(\bm p_{\bm n'},t')\rangle_A=V \delta_{\bm n,-\bm n'}
G^{(ij)}_{\mu\nu}(\bm p_{\bm n},t)\,,
\ee
with the usual infinite volume correlation function $G^{(ij)}_{\mu\nu}(\bm p,t)$, so that $\langle A^{(i)}(\bm x,t) A^{(j)}(\bm x',t')\rangle_A$ has the correct infinite volume limit (that is, independent of the volume $V$ as a local correlation function). Then, the one-quark Hamiltonian from $H_I$ becomes
\be
H_I(t)={g\over V}\sum_{\bm n,s}\sum_{\bm n',s'} {1\over \sqrt{2E_{p_{\bm n}}}}
 {1\over \sqrt{2E_{p_{\bm n'}}}} \bar u(\bm p_{\bm n},s)\gamma^\mu  u(\bm p_{\bm n'},s') A_\mu (\bm p_{\bm n}-\bm p_{\bm n'},t) a^\dagger_{\bm p_{\bm n},s} a_{\bm p_{\bm n'},s'}\,,\label{Hamiltonian}
\ee
where the color structure is omitted for notational simplicity. The {\it normalized} one-quark states are created by
\be
|\bm p_{\bm n},s\rangle \equiv a^\dagger_{\bm p_{\bm n},s}|0\rangle\,.
\ee

\begin{figure}[t]
 \centering
 \includegraphics[height=6cm,width=15cm]{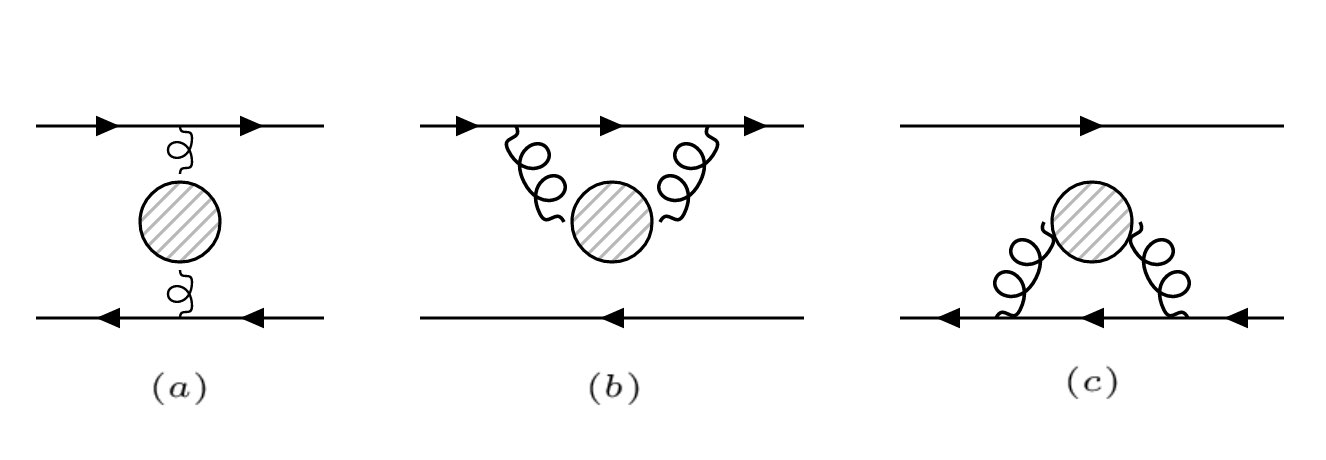}\caption{The ``cross" contribution (a), and the two self energy contributions (b) and (c). \label{fig2}}
 \end{figure}

To obtain the time evolution equation of the density matrix from (\ref{timeevol}) in perturbation
theory of $H_I$, we work in the interaction picture of $H_0$ and need to expand $U_{1,2}$ to quadratic order in $H_I$, since one-point functions of gluon fields vanish, $\langle A\rangle_A=0$, and the first non-vanishing correlation functions are the two-point functions.
Recall the interaction picture: $U(\Delta t,0)=U_0(\Delta t) U_I(\Delta t,0)$ where $\Delta t$ is the time step we are considering, $U_0(\Delta t)=e^{-i H_0 \Delta t}$ is the free evolution, and we set the initial time as $t_0=0$ without loss of generality. Then, we have the interaction picture evolution as
\be
U_I(\Delta t,0)\approx 1-i \int_0^{\Delta t}dt H_I^{\rm int}(t)+(-i)^2 \int_0^{\Delta t}dt \int_0^{t}dt'
H_I^{\rm int}(t)H_I^{\rm int}(t')+\cdots\,,
\ee
with $H^{\rm int}_I(t)=U_0(t)^\dagger H_I(t) U_0(t)$. Using these in (\ref{timeevol}), we have
\bear
\hat\rho(\Delta t)&=&U_0(\Delta t)\hat\rho(0)U_0^\dagger(\Delta t)+\int_0^{\Delta t} dt_1
\int_0^{\Delta t} dt_2 U_0(\Delta t)\langle H_I^{{\rm int}(1)}(t_1)\hat\rho(0) H_I^{{\rm int}(2)}(t_2)\rangle_A U_0^\dagger(\Delta t)\nonumber\\
&+& (-i)^2 U_0(\Delta t) \int_0^{\Delta t}dt_1\int_0^{t_1}dt_1' \langle H_I^{{\rm int}(1)}(t_1)H_I^{{\rm int}(1)}(t_1')\rangle_A \hat\rho(0) U_0^\dagger(\Delta t)\nonumber\\
&+& (+i)^2 U_0(\Delta t) \hat\rho(0) \int_0^{\Delta t}dt_2\int_0^{t_2}dt_2' \langle H_I^{{\rm int}(2)}(t_2')H_I^{{\rm int}(2)}(t_2)\rangle_A U_0^\dagger(\Delta t)\,,\label{evol}
\eear
where the second term in the first line comes from ``cross" combination obtained by one $H_I$ from $U_1$ and one from $U_2$, while the last two lines are ``self-energy" contributions coming from quadratic expansions in $H_I$ in each $U_1$ and $U_2$. We need the both types of contributions in order to make sure the probability conservation of the density matrix, that is,
${\rm Tr}(\hat\rho(\Delta t))={\rm Tr}(\hat\rho(0))$ to this order, that can be checked easily using the definition of correlation functions in Schwinger-Keldysh contours. See Figure \ref{fig2} for diagrammatic
representation of these contributions. 
The $H_I^{{\rm int}(1)}$ and $H_I^{{\rm int}(2)}$ are the Hamiltonians (\ref{Hamiltonian}) with $A^{(1)}$ and $A^{(2)}$ fields, respectively. As they are linear in $A$ fields, their correlation functions in (\ref{evol}) are proportional to the two-point correlation functions of background gluon fields, defined in (\ref{corr}).

Since our density matrix in (\ref{densityinp}) is diagonal in momentum space, and hence in energy spectrum of $H_0$, it commutes with $H_0$ and the first term in (\ref{evol}) is simply $\hat\rho(0)$.
Writing $\hat\rho(\bm p_{\bm n})$ in the helicity basis (\ref{helicitybasis}) in terms of $\rho_{s,s'}(\bm p_{\bm n},t)$ (now with time-dependence),
and using the explicit form of $H_I$ in (\ref{Hamiltonian}), the above evolution equation (\ref{evol}) can easily be translated to those of $\rho_{s,s'}(\bm p_{\bm n},t)$. From the identification $|\bm p_{\bm n},s\rangle=a^\dagger_{\bm p_{\bm n},s}|0\rangle$, it can be seen that the phase ambiguity of $\rho_{s,s'}(\bm p_{\bm n})$ via that of $|\bm p_{\bm n},s\rangle$ will cancel in the 
expression of the physical density matrix (\ref{densitym}), 
\be
\hat\rho(\bm p_{\bm n})=\sum_{s,s'}\xi_s(\bm p_{\bm n})\rho_{s,s'}(\bm p_{\bm n})\xi_{s'}^\dagger(\bm p_{\bm n})\,,
\ee
due to the fact that $u(\bm p_{\bm n},s)$ appearing in $H_I$ is proportional to $\xi_s(\bm p_{\bm n})$ and it shares the same phase with 
the $a^\dagger_{\bm p_{\bm n},s}$ (so that $H_I$ is unambiguous) and hence $|\bm p_{\bm n},s\rangle$.

When the time step $\Delta t$ of evolution is much larger than the correlation time of 
the gluon two-point functions, the cross and self energy terms in (\ref{evol}) will be linear in $\Delta t$, and the resulting evolution equation of density matrix will be of first order in time. 
In our leading log order, the dominant contribution to these terms come from the HTL contribution to the gluon two-point functions, with soft frequency-momenta in a range $gT\ll q\ll T$. This gives an estimate for the correlation time $\tau_c\lesssim (gT)^{-1}$ for leading log. As long as $\tau_c\ll\Delta t \ll 1/\delta E=\infty$, where $\delta E$ is the energy difference between the two quantum spin states, this is a valid description of the time evolution. 

More explicitly, for the cross term as an example, the two time integrals with the gluon two-point function can be written schematically as
\be
\int_0^{\Delta t} dt_1 \int_0^{\Delta t} dt_2 \,G^{(12)}(t_1-t_2)e^{i q^0 (t_1-t_2)}=\int_0^{\Delta t} dt_r\int_{-2t_r}^{2t_r} dt_a\, G^{(12)}(t_a)e^{iq^0 t_a}\,,
\ee
with a change of variable $t_r=(t_1+t_2)/2$ and $t_a=t_1-t_2$, and $q^0$ is some combination of energies of states (see the following expressions in this section). As $G^{(12)}(t_a)$ decays fast
beyond $t_a>\tau_c$, we can extend the range of $t_a$ integral to $[-\infty,+\infty]$ for most of $t_r$ values in $[0,\Delta t]$ when $\Delta t\gg\tau_c$: this gives the leading term linear in $\Delta t$. Then the above becomes $G^{(12)}(q^0)\Delta t$ where $G^{(12)}(q^0)$ is the Fourier transform of $G^{(12)}(t)$. A similar manipulation can be done for self energy terms to get the leading linear term in $\Delta t$.

After some algebra with these ingredients, and taking an infinite volume limit, we obtain a well-defined evolution equation for $\rho_{s,s'}(\bm p,t)$.
Writing 
\be
{d\over dt}\rho_{s,s'}(\bm p,t)= g^2 C_2(F)(\Gamma_{\rm cross}+\Gamma_{\rm self \,energy})\,,
\quad C_2(F)={N_c^2-1\over 2N_c}\,,
\ee
the cross contribution is given by
\be
\Gamma_{\rm cross}= \int {d^3 \bm p'\over (2\pi)^3}{1\over 4E_p E_{p'}}
\sum_{s'',s'''} [\bar u(\bm p,s)\gamma^\mu u(\bm p',s'')]\rho_{s'',s'''}(\bm p')[\bar u(\bm p',s''')\gamma^\nu u(\bm p,s')]G^{(12)}_{\mu\nu}(E_p-E_{p'},\bm p-\bm p')\label{cross}
\ee
and the self-energy contribution is a sum
\bear
\Gamma_{\rm self\,energy}=&-&\int {d^3\bm p'\over (2\pi)^3}{1\over 4E_p E_{p'}}\sum_{s'',s'''}[\bar u(\bm p,s)\gamma^\mu u(\bm p',s'')][\bar u(\bm p',s'')\gamma^\nu u(\bm p,s''')]\rho_{s''',s'}(\bm p)\nonumber\\
&\times&
\int^{+\infty}_0 dt_a\, G^{(11)}_{\mu\nu}(\bm p-\bm p',t_a) e^{i(E_p-E_{p'})t_a}\nonumber\\
&-&\int {d^3\bm p'\over (2\pi)^3}{1\over 4E_p E_{p'}}\sum_{s'',s'''}\rho_{s,s'''}(\bm p)[\bar u(\bm p,s''')\gamma^\mu u(\bm p',s'')][\bar u(\bm p',s'')\gamma^\nu u(\bm p,s')]\nonumber\\
&\times&
\int_{-\infty}^0 dt_a\, G^{(22)}_{\mu\nu}(\bm p-\bm p',t_a) e^{i(E_p-E_{p'})t_a}\,.\label{selfE}
\eear

In the Appendix, we prove the following important simplification in the self energy term, due to the rotational symmetry and P-even nature of thermal QCD background:
the $\bm p'$ integral appearing in the self energy term
\be
\int {d^3\bm p'\over (2\pi)^3}{1\over 2 E_{p'}}\sum_{s''}[\bar u(\bm p,s)\gamma^\mu u(\bm p',s'')][\bar u(\bm p',s'')\gamma^\nu u(\bm p,s''')] G^{(ij)}_{\mu\nu}(\bm p-\bm p',t_a)
\ee
is proportional to $\delta_{s,s'''}$, that is, it is non-zero only when the initial and final spins are the same, $s=s'''$, and moreover, the value doesn't depend on $s=\pm$. Physically, what it means is that the self energy can't flip the longitudinally polarized spin (i.e. helicity, $s/2$) due to rotational invariance of the background, and the self energy can't depend on the sign of helicity $s/2$ either, since $s$ flips under parity transformation. With this, the two terms in the self energy (\ref{selfE}) nicely combine 
to give
\bear
\Gamma_{\rm self\,energy}=&-&\int {d^3\bm p'\over (2\pi)^3}{1\over 4E_p E_{p'}}\sum_{s''}[\bar u(\bm p,s)\gamma^\mu u(\bm p',s'')][\bar u(\bm p',s'')\gamma^\nu u(\bm p,s)]\rho_{s,s'}(\bm p)\nonumber\\
&\times&
\left(\int^{+\infty}_0 dt_a\, G^{(11)}_{\mu\nu}(\bm p-\bm p',t_a)+\int_{-\infty}^0 dt_a\, G^{(22)}_{\mu\nu}(\bm p-\bm p',t_a)\right) e^{i(E_p-E_{p'})t_a}\nonumber\\
=&-&\int {d^3\bm p'\over (2\pi)^3}{1\over 4E_p E_{p'}}\sum_{s''}[\bar u(\bm p,s)\gamma^\mu u(\bm p',s'')][\bar u(\bm p',s'')\gamma^\nu u(\bm p,s)]\rho_{s,s'}(\bm p)\nonumber\\
&\times& G^{(21)}_{\mu\nu}(E_p-E_{p'},\bm p-\bm p')\,,\label{selfE2}
\eear
in terms of the Fourier transform of the correlation function $G^{(21)}_{\mu\nu}$, where we use an identity for Schwinger-Keldysh two-point functions, 
\be
G^{(11)}_{\mu\nu}(t)\theta(t)+G^{(22)}_{\mu\nu}(t)\theta(-t)=G^{(21)}_{\mu\nu}(t)\,,
\ee
to combine the two $t_a$ integrals. 

The appearance of $G^{(12)}$ in the cross term and $G^{(21)}$ in the self energy is a reflection of the generic feature that ensures the thermal detailed balance with the KMS relation (\ref{KMS}). In fact, using the identity 
\be
G^{(21)}=\left({n_B(q^0)+1\over n_B(q^0)}\right) G^{(12)}=e^{q^0/T} G^{(12)}\,,
\ee
one can easily check from (\ref{cross}) and (\ref{selfE2}) that the equilibrium thermal density matrix
\be
\rho^{\rm eq}_{s,s'}(\bm p)={z\over 2} \delta_{s,s'} e^{-E_p/T}\,,
\ee
with any fugacity constant $z=e^{\mu/T}$ makes the sum of $\Gamma_{\rm cross}$ and $\Gamma_{\rm self\, energy}$ terms vanishes. 
This equilibrium is also equivalently described by
\be
\hat\rho^{\rm eq}(\bm p)={z\over 2}e^{-E_p/T} {\bf 1}\,,\quad f^{\rm eq}(\bm p)=z e^{-E_p/T}\,,\quad \bm S^{\rm eq}(\bm p)=0\,.
\ee

The rest of the paper presents key elements of our computation of the integrals in (\ref{cross}) and (\ref{selfE2}) in leading log order. Readers who are interested in only the final results can go straight to section \ref{finalsec}.

\section{Leading log integrals with arbitrary quark mass \label{secLL}}

We first consider the evaluation of the cross term (\ref{cross}). Since the physical density matrix that is free of phase ambiguity that we discussed is $\hat\rho(\bm p)=\sum_{s,s'}\xi_s(\bm p)\rho(\bm p)_{s,s'}\xi_{s'}^\dagger(\bm p)$, we consider this object. Using also that
\be
\rho_{s,s'}(\bm p)=\xi_s^\dagger (\bm p)\hat\rho(\bm p)\xi_{s'}(\bm p)\,,
\ee
we can express (\ref{cross}) in terms of the unambiguous $\hat\rho(\bm p)$.
Since we will focus only on the soft $\bm q=\bm p-\bm p'\sim gT$ regime that produces the leading log, we change the integration variables from $\bm p'$ to $\bm q$,
\be
\int {d^3 \bm p'\over (2\pi)^3}=\int {d^3 \bm q\over (2\pi)^3}\,.
\ee
What needs to be computed in the resulting integrand is the following spinor summation, contracted with the gluon two-point function,
\be
\sum_{s,s',s'',s'''}\xi_s(\bm p)\bar u(\bm p,s)\gamma^\mu u(\bm p',s'')\xi^\dagger_{s''}(\bm p') \hat\rho(\bm p')\xi_{s'''}(\bm p') \bar u(\bm p',s''')\gamma^\nu u(\bm p,s')\xi^\dagger_{s'}(\bm p)G^{(12)}_{\mu\nu}(E_p-E_{p'},\bm q)\,,\label{spinorsum}
\ee
where $\bm p'=\bm p-\bm q$. 

We work in the Coulomb gauge, where the gluon two-point functions are written in terms of the longitudinal and transverse spectral densities, $\rho_L$ and $\rho_T$ respectively, (not to be confused with the density matrix)
\be
G_{\mu\nu}^{(12)}(q^0,\bm q)=n_B(q^0) \rho_{\mu\nu}(q^0,\bm q)\,,
\ee
with
\be
\rho_{\mu\nu}(q^0,\bm q)=
\left(\delta_{\mu 0}\delta_{\nu 0} \,\rho_L(q^0,q)+\Pi^T_{\mu\nu}(\bm q) \rho_T(q^0,q)\right)\,,
\ee
where $q\equiv |\bm q|$, and the transverse projection operator has only spatial components as
\be
\Pi^T_{ij}(\bm q)=(\delta_{ij}-\hat{\bm q}_i \hat{\bm q}_j)\,,\quad \hat{\bm q} \equiv \bm q/q\,.
\ee
From the explicit expression of spinor (\ref{spinor}), we have
\be
\bar u(\bm p,s)\gamma^\mu  u(\bm p',s'')=\left(\sqrt{(E_p-sp)(E_{p'}-s''p')} \xi_s^\dagger (\bm p)
\bar\sigma^\mu \xi_{s''}(\bm p')+\sqrt{(E_p+sp)(E_{p'}+s'' p'')}\xi_s^\dagger(\bm p)\sigma^\mu\xi_{s''}(\bm p')\right)\,,
\ee
where $\sigma^\mu=({\bf 1},\bm\sigma)$ and $\bar\sigma^\mu=({\bf 1},-\bm\sigma)$. Noting that 
\be \xi_s(\bm p)\xi^\dagger_s(\bm p)={\cal P}_s(\bm p)={1\over 2}\left({\bf 1}+s\hat{\bm p}\cdot\bm\sigma\right)\,,
\ee
 is the projection operator to the helicity $s/2$ state in spin space, the spinor sum in (\ref{spinorsum}) becomes a summation of various terms of the type,
 \be
 {\cal P}_s(\bm p) \sigma^\mu {\cal P}_{s''}(\bm p')\hat\rho(\bm p') {\cal P}_{s'''}(\bm p')\sigma^\nu {\cal P}_{s'}(\bm p)\,.
 \ee
 The computation of these spinor sum is challenging, but is doable with some efforts utilizing several properties of the projection operators, such as
 \be
 \sum_s {\cal P}_s(\bm p)={\bf 1}\,,\quad {\cal P}_s(\bm p) A{\cal P}_s(\bm p)={\rm Tr}\left(A {\cal P}_s(\bm p)\right) {\cal P}_s(\bm p)\,,
 \ee
 for any operator $A$.
 Note that  these expressions are free of phase ambiguity of $\xi_s(\bm p)$, due to the reasons as explained before.
 
 Since $G^{(12)}_{\mu\nu}(q^0,\bm q)=n_B(q^0)(\delta_{\mu 0}\delta_{\nu 0}\rho_L(q^0,q)+\Pi^T_{\mu\nu}(\bm q)\rho_T(q^0,q))$, we present our results of computation of (\ref{spinorsum}) 
in terms of longitudinal (that involves $\rho_L$) and transverse ($\rho_T$) gluon parts, respectively. The longitudinal part of (\ref{spinorsum}) is given as, omitting the obvious factor of $n_B(q^0)\rho_L(q^0,q)$ ($q^0\equiv E_p-E_{p'}$), and writing $\hat\rho(\bm p)$ in terms of the physical quantities (see (\ref{phys})), $\hat\rho(\bm p)={1\over 2}f(\bm p)+\bm S(\bm p)\cdot\bm\sigma$, 
\bear
&&{2\over (E_p+m)(E_{p'}+m)}\left(2(\bm p\cdot\bm p')(\bm p'\cdot\bm S(\bm p'))(\bm p\cdot\bm\sigma)-(p')^2(\bm p\cdot\bm S(\bm p'))(\bm p\cdot\bm\sigma)-p^2(\bm p'\cdot \bm S(\bm p'))(\bm p'\cdot\bm\sigma)   \right)\nonumber\\
&&+2\left((\bm p'\cdot\bm S(\bm p'))(\bm p\cdot\bm\sigma)-(\bm p\cdot \bm S(\bm p'))(\bm p'\cdot\bm\sigma)\right)+2\left(E_p E_{p'}+\bm p\cdot\bm p'+m^2\right)\hat\rho(\bm p')\,.
\eear
The transverse part of (\ref{spinorsum}), that is proportional to $n_B(q^0)\rho_T(q^0, q)$, is
obtained after a long computation as
\bear
&&2\left(E_p E_{p'}-(\bm p\cdot\hat{\bm q})(\bm p'\cdot\hat{\bm q})-m^2\right) f(\bm p')-4{(E_p+m)\over (E_{p'}+m)}(\bm p'\cdot\bm S(\bm p'))(\bm p'\cdot\hat{\bm q})(\hat{\bm q}\cdot\bm\sigma)
\nonumber\\&+&4\left(-{(E_{p'}+m)\over (E_p+m)}(\hat{\bm q}\cdot\bm S(\bm p'))(\bm p\cdot\hat{\bm q})+\bm p'\cdot\bm S(\bm p')\right)(\bm p\cdot\bm\sigma)\nonumber\\
&+&4(E_p E_{p'} -m^2)(\hat{\bm q}\cdot\bm S(\bm p'))(\hat{\bm q}\cdot\bm\sigma)
+4 \left((\bm p'\times\bm\hat{\bm q})\cdot\bm S(\bm p')\right)\left((\bm p\times\hat{\bm q})\cdot\bm\sigma\right)\,.
\eear

We have checked the validity of the above results at least in two special limits: 1) $\bm p'=\bm p$ limit (treating $\hat{\bm q}$ arbitrary), and 2) massless ($m=0$) limit. In both limits, the spinor sum (\ref{spinorsum}) reduces to $s''=s$ and $s'''=s'$ cases only, due to the fact that $\bar u(\bm p,s)\gamma^\mu u(\bm p',s'')$ vanishes in these limits unless $s=s''$, which can be easily checked from the explicit spinor expression (\ref{spinor}). Using this fact, one can compute (\ref{spinorsum}) in these limits directly, and then can compare with the above results in the same limits. The longitudinal part is easy to compare, but the comparison of the transverse part needs some non-trivial identities.
In the limit 1), one needs the following ``dyad" identity ($\otimes$ is a dyad product of row and column vectors)
\be
\hat{\bm p}\otimes\hat{\bm p}-(\hat{\bm p}\cdot\hat{\bm q})\hat{\bm q}\otimes\hat{\bm p}-(\hat{\bm p}\cdot\hat{\bm q})\hat{\bm p}\otimes\hat{\bm q}-\hat{\bm q}\otimes\hat{\bm q}
+(\hat{\bm p}\times\hat{\bm q})\otimes (\hat{\bm p}\times \hat{\bm q})=(1-(\hat{\bm p}\cdot\hat{\bm q})^2) {\bm I}\,,
\ee 
for any two unit vectors $\hat{\bm p}$ and $\hat{\bm q}$,
where $\bm I$ is the $3\times 3$ identity dyad (matrix). In the case of the limit 2), one needs a more non-trivial identity that we checked by Mathematica,
\bear
&&(1-\hat{\bm p}\cdot \hat{\bm p'})\hat{\bm q}\otimes \hat{\bm q}-(\hat{\bm p}\cdot\hat{\bm q})\hat{\bm q}\otimes(\hat{\bm p}-\hat{\bm p}')+\hat{\bm p}'\otimes \hat{\bm p}-\hat{\bm p}\otimes\hat{\bm p}'+(\hat{\bm q}\cdot\hat{\bm p}')(\hat{\bm p}-\hat{\bm p}')\otimes\hat{\bm q}\nonumber\\
&+&(\hat{\bm p}\cdot\hat{\bm p}'-(\hat{\bm p}\cdot\hat{\bm q})(\hat{\bm p}'\cdot\hat{\bm q})){\bm I}
= -(\hat{\bm p}\cdot\hat{\bm q})\hat{\bm q}\otimes\hat{\bm p}+\hat{\bm p}'\otimes\hat{\bm p}
-(\hat{\bm p}'\cdot\hat{\bm q})\hat{\bm p}'\otimes \hat{\bm q}+\hat{\bm q}\otimes \hat{\bm q}+(\hat{\bm p}'\times \hat{\bm q})\otimes (\hat{\bm p}\times\hat{\bm q})\,,\nonumber\\
\eear
for any three unit vectors $\hat{\bm p}$, $\hat{\bm p}'$ and $\hat{\bm q}$. These agreements give us confidence on the validity of the above spinor sum results.
 
 The computation of spin sum in the self energy (\ref{selfE2}) is simpler.
 First note that the self energy term has a simple structure
 \be
 \Gamma_{\rm self\,energy}=-\gamma \,\hat\rho(\bm p)\,,\label{selfenergy}
 \ee
that is, it is a constant ($\gamma$) times of the identity operator in both the spin and momentum space. The ``damping rate" $\gamma$ is (recall $\bm p'=\bm p-\bm q$)
\bear
\gamma=\int {d^3\bm q\over (2\pi)^3}{1\over 4E_p E_{p'}}\sum_{s''}[\bar u(\bm p,s)\gamma^\mu u(\bm p',s'')][\bar u(\bm p',s'')\gamma^\nu u(\bm p,s)]G^{(21)}_{\mu\nu}(E_p-E_{p'},\bm q)\,.
\eear
Recalling that this expression doesn't depend on $s$ (due to parity invariance as proved in Appendix), it turns out to be easier to compute the spin sum by expressing it as
\be
\gamma={1\over 2}\int {d^3\bm q\over (2\pi)^3}{1\over 4E_p E_{p'}}\sum_{s,s''}[\bar u(\bm p,s)\gamma^\mu u(\bm p',s'')][\bar u(\bm p',s'')\gamma^\nu u(\bm p,s)]G^{(21)}_{\mu\nu}(E_p-E_{p'},\bm q)\,,\label{gam}
\ee
removing any reference to $s$. For the longitudinal gluon contribution ($\mu=\nu=0$), the spin sum becomes (recall $G^{(21)}_{\mu\nu}(q^0,\bm q)=(n_B(q^0)+1)(\delta_{\mu 0}\delta_{\nu 0}\rho_L(q^0,q)+\Pi^T_{\mu\nu}(\bm q)\rho_T(q^0,q))$)
\be
{1\over 2}\sum_{s,s''}[\bar u(\bm p,s)\gamma^0 u(\bm p',s'')][\bar u(\bm p',s'')\gamma^0 u(\bm p,s)]=2\left(E_p E_{p'}+\bm p\cdot\bm p'+m^2\right)\,,
\ee
and for the transverse gluon contribution, we obtain
\be
{1\over 2}\sum_{s,s''}[\bar u(\bm p,s)\gamma^i u(\bm p',s'')][\bar u(\bm p',s'')\gamma^j u(\bm p,s)]\Pi^T_{ij}(\bm q)=4\left(E_p E_{p'}-(\bm p\cdot\hat{\bm q})(\bm p'\cdot\hat{\bm q})-m^2\right)\,.
\ee

After (\ref{spinorsum}) and the spin sum in (\ref{gam}) is computed, what remains is to compute  
the $\bm q$ integrations in (\ref{cross}) and (\ref{gam}) to leading log order, with the gluon spectral densities $\rho_{L/T}(q^0,q)$ given by the well-known HTL contributions.
For completeness, they are given by
\be
\rho_L(q^0,q)=-2\,{\rm Im}\left(1\over q^2-\Pi_L\right)\,,\quad
\Pi_L=-m_D^2 \left(1+(q^0/2q)\log\left(q^0-q+i\epsilon\over q^0+q+i\epsilon\right)\right)\,,
 \label{spectL}
\ee
where $m_D\sim gT$ is the Debye mass, and
\be
\rho_T(q^0,q)=  2\,{\rm Im}\left(1\over q^2-(q^0)^2-\Pi_T\right)\,,\label{spectT}
\ee
where
\be
\Pi_T=-{m_D^2\over 2}\left((q^0/q)^2+\left((q^0/q)^2-1\right)(q^0/2q)\log\left(q^0-q+i\epsilon\over
q^0+q+i\epsilon\right)\right)
\,.
\ee
As in the computations of shear viscosity and conductivities in massless limit \cite{Arnold:2000dr}, as well as in the computation of 
diffusion constant of heavy quark limit \cite{Moore:2004tg}, we find that the leading log contribution comes from the same soft $q$ regime, for arbitrary quark mass, where the log arises from the range $gT\ll q\ll T$.
Physically, this contribution represents the t-channel scatterings with background thermal particles of hard scale ($p\sim T$) with a soft gluon exchange of momentum $(q^0,\bm q)$.
We emphasize that these HTL contributions include only the thermal background gluons and light quarks, which means that we don't include the scatterings with the other massive quarks present 
in the background plasma. This is the same ``diluteness" assumption we explained at the beginning of section \ref{sec2}. The Debye mass in this case is given by
\be
m_D^2={g^2 T^2\over 6}(2N_c+N_F)\,,
\ee
where $N_F=2$ is the number of light flavors.

We follow the known steps of computing $\bm q$ integration in leading log order \cite{Arnold:2000dr,ValleBasagoiti:2002ir,Aarts:2002tn,Hong:2010at,Jimenez-Alba:2015bia}.
A first step in this $\bm q$ integration is to make a change of variable from the azimuthal angle $\cos\theta_{pq}$ between $\bm p$ and $\bm q$ to the energy transfer $q^0=E_{p}-E_{p'}$ (recall $\bm p'\equiv\bm p-\bm q$), where they are related by
\be
q\cos\theta_{pq}=\hat{\bm p}\cdot\bm q={E_p\over p}q^0+{q^2-(q^0)^2\over 2p}\equiv q_L\,.
\ee
The $q^0$ has a maximum $q^0_{\rm max}$ (minimum $q^0_{\rm min}$) when $\theta_{pq}=0$ ($\pi$), and
\be
q^0_{\rm max/min}=\sqrt{p^2+m^2}-\sqrt{(p\mp q)^2+m^2}\approx \pm{p\over E_p} q-{m^2 q^2\over 2 E_p^3}+{\cal O}(q^3)\,.\label{range}
\ee
Note that $q^0_{\rm min}$ is different from $-q^0_{\rm max}$ by a term proportional to $q^2$,
that is present only in the massive case. We will see that this $q^2$ correction to the $q^0$ integration range, that is absent in massless case, gives rise to the same leading log contribution to the final result,
so it is important to keep it to this order.
From this, we can convert $\bm q$ integration in (\ref{cross}) and (\ref{gam}) into an integration of two variables $(q^0,q)$, 
\bear
&&\int {d^3\bm q\over (2\pi)^3} {1\over 2 E_{p'}}\, ({\rm spinor\, sum}) \,\rho_{L/T}(E_{p}-E_{p'},q)\nonumber\\&=&
{1\over 2p}\int_0^\infty {dq\,q\over (2\pi)} \int_{q^0_{\rm min}}^{q^0_{\rm max}}{dq^0\over (2\pi)} \,({\rm spinor\, sum})\,\rho_{L/T}(q^0,q)\bigg|_{\hat{\bm p}\cdot\bm q\to {E_p\over p}q^0+{q^2-(q^0)^2\over 2p}}\,,\label{qint}
\eear
where (spinor sum) is the spinor part that we computed above, and the integration of polar angle around $\hat{\bm p}$ axis gives $(2\pi)$, after we make the spinor part to be independent of polar angle, exploiting rotational symmetry, since $\rho_{L/T}(q^0,q)$ depends on $\bm q$ only via $q=|\bm q|$. Specifically, we can replace
\be
\bm q^i \to q_L\hat{\bm p}^i=(\hat{\bm p}\cdot\bm q)\hat{\bm p}^i\,,\quad \bm q^i\bm q^j\to
q_L^2\hat{\bm p}^i\hat{\bm p}^j+{1\over 2}(\delta^{ij}-\hat{\bm p}^i\hat{\bm p}^j) (q^2-q_L^2)\,.
\ee

One then computes $(q^0,q)$ integration with the gluon spectral densities $\rho_{L/T}(q^0,q)$
to the desired order that produces the leading log in the final result. 
Since the leading log comes from soft $(q^0,q)\sim gT$ regime, one expands the (spinor sum) part in power series of soft $(q^0,q)\ll (p,E_p,m)\sim T$: it is sufficient to keep only up to linear order in $(q^0,q)$, as higher powers give higher order terms in $g$. The total sum of the cross and self energy terms, (\ref{cross}) and (\ref{selfenergy}) respectively, takes a form
of (\ref{qint}) after this expansion, where the (spinor sum) part has a structure of
\be
({\rm spinor\,sum})=C^0(q^0/q)+q^0C^1(q^0/q)+{\cal O}(q^2)\,,
\ee
with two functions $C^{0,1}(q^0/q)$ on $q^0/q$.
There are two important features in this result, that makes the leading log contribution possible:
1) In principle, since the (spinor sum) contains $n_B(q^0)\sim T/q^0$ for $q^0\sim gT\ll T$ (see (\ref{KMS})), the expansion could start from $(1/q^0) C^{-1}(q^0/q)$, instead of $C^0(q^0/q)$.
In fact, both the cross and the self energy terms start from this order, but their sum cancels to this order. If this cancellation was absent, the final result of spin evolution rate would have been dominated by this order, which gives $g^2\log(1/g)$, instead of $g^4\log(1/g)$ that we find.
This contribution would come from the ultra-soft range $g^2 T\ll q\ll gT$, which represents ``small angle" scatterings, contrary to our range $gT\ll q\ll T$ for $g^4\log(1/g)$ that represents ``large angle" scatterings. This cancellation is important also in computations of shear viscosity and charge conductivities  (but not in ``color" conductivity \cite{Arnold:1996dy,MartinezResco:2000pz}), and has been shown to be related to conservation Ward identities of energy-momentum and charge currents \cite{Aarts:2002tn}. The same cancellation we observe in our spin density matrix suggests it may be related to angular momentum conservation. This cancellation also adds confidence that our computation of spin sum is correct. 2) In principle, the functions $C^{0,1}(q^0/q)$ could be any function on $q^0/q$, but they turn out to be even functions on $q^0/q$, which is crucial to have the final $g^4\log(1/g)$ rate.
This feature is important due to the fact that the spectral densities, $\rho_{L/T}$, are odd functions on $q^0$.
If $q^0_{\rm min}$ was precisely equal to $-q^0_{\rm max}$ (as in the massless case), the $q^0$ integral of $C^0(q^0/q)$ would have vanished, and the first non-vanishing result would come from the next order $q^0 C^1(q^0/q)$ term, which gives $g^4\log(1/g)$ rate. Due to the $q^2$ correction to the $q^0$ integration range, $q^0_{\rm max/min}$, in our massive case (see (\ref{range})), the $C^0(q^0/q)$ integral does contribute, but since this correction is one higher order than the leading range, the result is of the same order as the one from $q^0 C^1(q^0/q)$, that is, the same $g^4\log(1/g)$ rate.

Let us define $q^0$ integrals of spectral densities, that are needed in our computation described above,
\be
J^{L/T}_n=\int_{q^0_{\rm min}}^{q^0_{\rm max}} {dq^0\over (2\pi)} (q^0)^{2n-1} \rho_{L/T}(q^0,q)\,.
\ee
In the massless case, only integer $n$ survives due to $q_{\rm min}=-q_{\rm max}$ in that case, and they can be computed by a sum-rule technique \cite{ValleBasagoiti:2002ir,Aarts:2002tn,Jimenez-Alba:2015bia}, utilizing analytic property of the spectral densities. In our massive case, and only for the range of our interests, $gT\ll q\ll T$,
the spectral densities can be simplified to produce the results for $J_n^{L/T}$ to our desired order
\be
\rho_L(q^0,q)\approx {\pi m_D^2 q^0\over q^5}\,,\quad
\rho_T(q^0,q)\approx {\pi m_D^2 \left(1-(q^0/q)^2\right)(q^0/2q)\over \left(q^2-(q^0)^2+m_D^2/2\right)^2}\,,
\ee
where $m_D\sim gT$ is the Debye mass.
One can easily check that in the massless limit, these spectral densities produce the same results for $J_n^{L/T}$ from the sum-rule technique. These expressions are obtained from the full expressions, (\ref{spectL}) and (\ref{spectT}), by using the hierarchy $m_D\sim gT \ll (q^0,q)\ll T$.
It is convenient to write $J^{L/T}_n$ as
\be
J^{L/T}_n={m_D^2\over q^{(4-2n)}}j^{L/T}_n\quad (n={\rm integer})\,,\quad J^{L/T}_n={m_D^2\over q^{(3-2n)}}{m^2\over E_p^3}j^{L/T}_n\quad (n={\rm half\,integer})\,,
\ee
in terms of the dimensionless coefficient functions $j^{L/T}_n$ on $(p,E_p,m)$, after extracting the dependence on $q$ explicitly as above.
By explicit evaluations, we find them as in Table 1.
\begin{table}[h]
\centering
\begin{tabular}{ |p{4cm}|p{4cm}|  }
\hline
 $j^L_{0} = \frac{p}{E_p}$ & $j^T_0 = \frac{\eta_p}{2}$   \\ 
 \hline
 $j^L_{1/2} = -\frac{p}{2E_p}$  & $j^T_{1/2} = -\frac{pE_p}{4m^2}$   \\
 \hline
 $j^L_{1} = \frac{p^3}{3E^3_p}$ & $ j^T_1 = \frac{\eta_p}{2}-\frac{p}{2E_p}$   \\
 \hline
 $j^L_{3/2}=-\frac{2p^3}{E^3_p}$ & $j^T_{3/2} = -\frac{p^3}{4 m^2 E_p}$   \\ 
 \hline 
 $j^L_2 = \frac{p^5}{5E^5_p}$  & $j^T_2 =  \frac{\eta_p}{2}-\frac{p}{2E_p} - \frac{p^3}{6E^3_p} $ \\ 
 \hline
\end{tabular}\caption{The coefficient functions $j^{L/T}_n$ ($\eta_p = \frac{1}{2}\ln \frac{E_p + p}{E_p - p}$ is rapidity).}
\end{table} 
Note that the half-integer $n$ cases are needed for the contributions from $C^0(q^0/q)$ as we explained above, which exist only in the massive case.
After doing $q^0$ integration using these formula, one finally performs $q$ integration in (\ref{qint}) to get the leading log result, where $q$ ranges in $m_D\ll q\ll T$: these boundaries come from the fact that our expression for the integrand is valid only in this range. The log arises from
\be
\int_{m_D}^T {dq\over q}\sim \log(T/m_D)\sim \log(1/g)\,.
\ee

\section{Quantum kinetic equation for spin polarization of massive quarks \label{finalsec} }

After a lengthy, but straightforward computation that we describe in the previous section, we present our final result for the time-evolution of the spin density matrix in momentum space, $\hat\rho(\bm p)={1\over 2}f(\bm p)+\bm S(\bm p)\cdot\bm\sigma$, in leading log order of $g^4\log(1/g)$. We write these evolution equations as
\be
{\partial f(\bm p,t)\over\partial t}=C_2(F) {m_D^2 g^2\log(1/g)\over (4\pi)}\,{1\over 2p E_p}\,\Gamma_f\,,\quad
{\partial \bm S(\bm p,t)\over\partial t}=C_2(F){m_D^2 g^2\log(1/g)\over (4\pi)}\,{1\over 2p E_p}\,\bm\Gamma_S\,,
\ee
where $\Gamma_f$ and $\bm\Gamma_S$ are diffusion-like differential operators in momentum space, that contain up to second order derivatives in $\bm p$. The $\Gamma_f$ is given by ($\bm\nabla_p\equiv\partial/\partial\bm p$ and $(\bm p\cdot\bm\nabla_p)^2 f\equiv \bm p\cdot\bm\nabla_p\left(\bm p\cdot\bm\nabla_p f\right)$)
\bear
\Gamma_f&=& 4\left(-{m^2\over E_p}j^L_{1/2}+E_p j^L_1-{m^2\over E_p^3}(p^2 j^T_{1/2}-E_p^2 j^T_{3/2}) +E_p(j^T_1-j^T_2)\right) f(\bm p)\nonumber\\
&+& \left(E_p^2 T\left(j^L_0-{E_p^2\over p^2} j^L_1\right)+Tp^2 j^T_0-2TE_p^2 j^T_1+{TE_p^4\over p^2}j^T_2   \right)\bm\nabla_p^2 f(\bm p)\nonumber\\
&+& \left( -{TE_p^2\over p^2}\left(j^L_0-{3 E_p^2\over p^2} j^L_1\right) -Tj^T_0+{4 TE_p^2\over p^2}j^T_1-{3 TE_p^4\over p^4}j^T_2  \right) (\bm p\cdot\bm\nabla_p)^2 f(\bm p)
\nonumber\\
&+& {1\over p^2} \Bigg(-{4m^2 T}j^L_{1/2}-{T E_p^2}j^L_0+\left({6TE_p^2}+{2E_p^3}-{3T E_p^4\over p^2}\right)j^L_1-{4m^2T\over  E_p^2}(p^2 j^T_{1/2}-E_p^2 j^T_{3/2})\nonumber\\
&& -Tp^2j^T_0+2\left(p^2E_p+Tp^2+{TE_p^2}\right)j^T_1+{E_p^2}\left(-2E_p-6T+{3TE_p^2\over p^2}\right)j^T_2\Bigg) ({\bm p}\cdot\bm\nabla_p)f(\bm p)\,.\nonumber\\\label{finalF}
\eear
This result passes a very non-trivial test of the expected detailed balance: one can check that $\Gamma_f=0$ when $f(\bm p)=f^{\rm eq}(\bm p)=z e^{-E_p/T}$ for any constant $z$.
It should be emphasized that this check is satisfied irrespective of the values of $j^{L/T}_n$,
because the detailed balance is independent of details of the spectral densities that determine $j^{L/T}_n$. This gives us confidence that our computation is correct. We also note that our result for $\Gamma_f$ that provides the local collision term in leading log order, together with free streaming advection term in Boltzmann equation, can be used to compute several conventional transport coefficients, such as shear viscosity and electric conductivity, arising from dilute massive quarks.

For the spin polarization part, we obtain ($i=1,2,3$ denotes a spatial index for vector)
\bear
\bm\Gamma_S^i&=& \Bigg(-{4m^2\over E_p}j^L_{1/2}+Tj^L_0+\left(4E_p-{TE_p^2\over p^2}\right) j^L_1-{4m^2\over E_p^3}(p^2 j^T_{1/2}-E_p^2 j^T_{3/2}) +Tj^T_0\nonumber\\
&& +\left(4E_p+T-{3TE_p^2\over p^2}\right)j^T_1+\left(-4E_p+{TE_p^2\over p^2}\right)j^T_2 \Bigg) \bm S^i(\bm p)\nonumber\\
&+& \left(E_p^2 T\left(j^L_0-{E_p^2\over p^2} j^L_1\right)+Tp^2 j^T_0-2TE_p^2 j^T_1+{TE_p^4\over p^2}j^T_2   \right)\bm\nabla_p^2 \bm S^i(\bm p)\nonumber\\
&+& \left( -{TE_p^2\over p^2}\left(j^L_0-{3 E_p^2\over p^2} j^L_1\right) -Tj^T_0+{4 TE_p^2\over p^2}j^T_1-{3 TE_p^4\over p^4}j^T_2  \right) (\bm p\cdot\bm\nabla_p)^2 \bm S^i(\bm p)
\nonumber\\
&+&{1\over p^2} \Bigg(-{4m^2 T}j^L_{1/2}-{T E_p^2}j^L_0+\left({6TE_p^2}+{2E_p^3}-{3T E_p^4\over p^2}\right)j^L_1-{4m^2T\over  E_p^2}(p^2 j^T_{1/2}-E_p^2 j^T_{3/2})\nonumber\\
&& -Tp^2j^T_0+2\left(p^2E_p+Tp^2+{TE_p^2}\right)j^T_1+{E_p^2}\left(-2E_p-6T+{3TE_p^2\over p^2}\right)j^T_2\Bigg) (\bm p\cdot\bm\nabla_p)\bm S^i(\bm p)\nonumber\\
&+&2T\left({E_p\over E_p+m}\left(j^L_0-{E_p^2\over p^2}j^L_1\right)+j^T_0-{E_p\over p^2}(2E_p-m)j^T_1+{E_p^3\over p^2(E_p+m)}j^T_2\right)\bm p^i(\bm\nabla_p\cdot\bm S(\bm p))\nonumber\\
&-&2T\left({E_p\over E_p+m}\left(j^L_0-{E_p^2\over p^2}j^L_1\right)+j^T_0-{E_p\over p^2}(2E_p-m)j^T_1+{E_p^3\over p^2(E_p+m)}j^T_2\right)\bm\nabla_p^i(\bm p\cdot\bm S(\bm p))\nonumber\\
&-&{T\over p^2}\Bigg({E_p-m\over E_p+m}\left(j^L_0-{E_p^2\over p^2}j^L_1\right)+j^T_0-\left(1+{3E_p^2\over p^2}-{2E_p\over E_p+m}\right)j^T_1\nonumber\\ && +{E_p^2\over p^2}\left(3+{4p^2\over (E_p+m)^2}-{6E_p\over E_p+m}\right)j^T_2\Bigg)\bm p^i ({\bm p}\cdot\bm S(\bm p))\,.\label{finalS}
\eear
Note that $f(\bm p)$ and $\bm S(\bm p)$ do not mix with each other in these equations.

There is a highly non-trivial test of the above result in the massless limit.
Note that our computation doesn't include the quark-gluon conversion processes that becomes of the same order in the massless limit, and also our values of $J^{L/T}_n$ do not have the correct massless limit, so the massless limit of the above result should not be taken as the true result for the massless case.
What we are testing is the ``consistency'' of the above equations with the ``chirality conservation" in the massless limit, and this test is a {\it kinematical} one, and should hold true for each scattering processes included, independent of details of spectral densities, that is, the values of $J^{L/T}_n$.  In the massless limit, the negative helicity state ($s=-1$, left-handed) and positive helicity state ($s=+1$, right-handed) are decoupled, and do not mix by gauge interactions. 
The spin density matrix should then take the following decoupled form
\be
\hat\rho(\bm p)=f_+(\bm p){\cal P}_+(\bm p)+f_-(\bm p){\cal P}_-(\bm p)\,,\quad {\cal P}_{\pm}(\bm p)={1\over 2}\left({\bf 1}\pm \hat{\bm p}\cdot\bm\sigma\right)\,,
\ee
as a sum of positive and negative helicity chiral quark contributions, where ${\cal P}_\pm$ are nothing but the spin projection operators to the two decoupled helicity states.
The $f_\pm(\bm p)$ are the number distribution functions of chiral quarks of helicity $s/2=\pm1/2$
in momentum space. In a parity-even background that we are considering, $f_+$ and $f_-$ should satisfy the same evolution equation. Writing the above density matrix as
\be
\hat\rho(\bm p)={1\over 2}\left(f_+(\bm p)+f_-(\bm p)\right)+{1\over 2}\left(f_+(\bm p)-f_-(\bm p)\right)\hat{\bm p}\cdot\bm\sigma\,,
\ee
we see the correspondence to our variables $f(\bm p)$ and $\bm S(\bm p)$ as, 
\be
f(\bm p)=f_+(\bm p)+f_-(\bm p)\,,\quad \bm S={1\over 2}\left(f_+(\bm p)-f_-(\bm p)\right)\hat{\bm p}\equiv f_s(\bm p)\hat{\bm p}\,.
\ee
Since $f_+$ and $f_-$ satisfy the same evolution equation, the two functions $f(\bm p)$ and $f_s(\bm p)$ should satisfy the same equation as well. This means that our above result, when we take the massless limit while keeping $J^{L/T}_n$ arbitrary, should pass the following non-trivial tests: 1) the evolution equation for $\bm S(\bm p)$ must admit a consistent Ansatz, $\bm S(\bm p)=f_s(\bm p)\hat{\bm p}$, and 2) the resulting evolution equation for $f_s(\bm p)$ must be the same as the one for $f(\bm p)$ in the massless limit. Both tests require non-trivial cancellations between various terms in (\ref{finalS}), and it is amusing to check that the tests are satisfied by our results (\ref{finalF}) and (\ref{finalS}):
the $f({\bm p})$ and $f_s(\bm p)$ satisfy the same evolution equation in the massless limit with
\bear
\Gamma_f^{m=0}
&=& 4p\left(j^L_1+j^T_1-j^T_2)\right) f(\bm p)+ p^2 T\left(j^L_0-j^L_1+j^T_0-2j^T_1+j^T_2   \right)\bm\nabla_p^2 f(\bm p)\nonumber\\
&+& T\left( -j^L_0+3  j^L_1 -j^T_0+4 j^T_1-3 j^T_2  \right) (\bm p\cdot\bm\nabla_p)^2 f(\bm p)
\nonumber\\
&+& \left(-{T}j^L_0+(3T+2p)j^L_1-j^T_0+2(p+2T)j^T_1-(3T+2p)j^T_2\right) ({\bm p}\cdot\bm\nabla_p)f(\bm p)\,.\nonumber\\\label{finalF2}
\eear
This also means that the massless limit allows a broader set of equilibria as
\be
\hat\rho^{\rm eq}(\bm p)=z_+ e^{-p/T} {\cal P}_+(\bm p)+z_- e^{-p/T} {\cal P}_-(\bm p)\,,
\ee
with arbitrary chiral fugacity constants $z_\pm=e^{\mu_\pm/T}$. With these remarkable checks satisfied, we become confident that the results (\ref{finalF}) and (\ref{finalS}) are correct.

Using the explicit values of $j^{L/T}_n$ given in Table 1, we have the following expression for the $\Gamma_f$,
\bear
\Gamma_f &=& 2 p f(\bm p) + \left(\frac{3}{2}T E_p p  - \frac{T E^3_p}{2p} + \frac{\eta_p T m^4}{2p^2}\right)\bm\nabla_p^2 f(\bm p) + \frac{Tm^2}{2p^2}\left(\eta_p + \frac{3E_p}{p} - \eta_p \frac{3E^2_p}{p^2}\right)(\bm p\cdot\bm\nabla_p)^2 f(\bm p) \nonumber\\
&+& \frac{1}{p^2}\left(p E^2_p - \frac{\eta_p T m^2}{2} - \eta_p E_p m^2 -\frac{3TE_p m^2}{2p} + \frac{3\eta_p T m^2 E^2_p}{2p^2}\right)({\bm p}\cdot\bm\nabla_p)f(\bm p)\,.
\eear
It can be checked again that the detailed balance condition is satisfied with $f^{\rm eq}(\bm p)=z e^{-E_p/T}$. For $\bm \Gamma^i_S$, we have
\bear
\bm \Gamma^i_S &=& \left(2p + \frac{TE_p}{p} - \frac{ \eta_p m^2 T}{p^2}\right)\bm S^i(p) + \left(p T E_p - \frac{m^2 T E_p}{2p} + \frac{\eta_p m^4 T}{2p^2}\right)\bm\nabla_p^2 \bm S^i(\bm p) \nonumber\\
&+& \left(\frac{\eta_p m^2 T}{2p^2}\left(1-\frac{3E^2_p}{p^2}\right) + \frac{3m^2T E_p}{2p^3}\right)(\bm p\cdot\bm\nabla_p)^2\bm S^i(\bm p)\nonumber \\
&+&\frac{1}{p^2}\left(pE^2_p -\frac{3m^2TE_p}{2p} + \eta_p m^2\left(-E_p -\frac{T}{2}+\frac{3TE^2_p}{2p^2}\right)\right)(\bm p\cdot\bm\nabla_p)\bm S^i(\bm p)\nonumber \\
&+&2T \left(\eta_p\left(\frac{1}{2} -\frac{E^2_p}{p^2} + \frac{mE_p}{2p^2} + \frac{E^3_p}{2p^2(E_p + m)}\right) + \frac{E_p}{p} -\frac{m}{2p} -\frac{m^2}{2p(E_p + m)}\right)\bm p^i(\bm\nabla_p\cdot\bm S(\bm p))\nonumber \\
&-&2T\left(\eta_p\left(\frac{1}{2} -\frac{E^2_p}{p^2} + \frac{mE_p}{2p^2} + \frac{E^3_p}{2p^2(E_p + m)}\right) + \frac{E_p}{p} -\frac{m}{2p} -\frac{m^2}{2p(E_p + m)}\right)\bm\nabla_p^i(\bm p\cdot\bm S(\bm p)) \nonumber\\
&-&\frac{T}{p^2}\left(\frac{E_p(E_p + 2m)}{p(E_p + m)} + \frac{\eta_p m E_p}{E_p+m}\left(-\frac{3E_p}{p^2}+\frac{1}{E_p +m}\right)\right)\bm p^i ({\bm p}\cdot\bm S(\bm p))\,.
\eear

\section{Discussion}

Our work is a small step toward a more complete picture of quantum kinetic theory of spin dynamics in perturbative QCD plasma.
It is important to extend our work, going beyond the spatial homogeneous limit.
This would introduce a spin density matrix $\hat\rho(\bm x,\bm p)$ that depends on both position and momentum, or equivalently $\hat\rho(\bm p_1,\bm p_2)$ which is non-diagonal in momentum space, as explained in the introduction. One could expect that the resulting quantum kinetic equation would look like
\be
\left({\partial\over\partial t}+\bm v_p\cdot {\partial\over \partial\bm x}\right)\hat\rho(\bm x,\bm p)=\Gamma\cdot  \hat\rho(\bm x,\bm p)\,,\quad \bm v_p\equiv {\bm p\over E_p}\,,
\ee
where $\Gamma$, the quantum kinetic collision term, is what we compute in this work.
As discussed in the introduction, this picture has to be improved by including 1) gradient corrections to the collision term $\Gamma$ (that is, corrections that involve $\partial_{\bm x}\hat\rho$), in order to allow spin-orbital angular momentum exchange, and 2) the effects of background electromagnetic fields in both free streaming and collision terms. The 2) for the free streaming part was recently studied in Ref.\cite{Mueller:2019gjj,Weickgenannt:2019dks,Gao:2019znl,Hattori:2019ahi} for massive quarks, extending the recent development of chiral kinetic theory for massless chiral quarks \cite{Son:2012wh,Stephanov:2012ki,Chen:2012ca,Duval:2014ppa,Mueller:2017arw,Gao:2017gfq,Carignano:2018gqt}, for which a Berry's curvature in momentum space due to spin projection plays a critical role. The 1) is also expected to be intimately related to the ``side-jump'' phenomenon in chiral kinetic theory \cite{Chen:2015gta,Hidaka:2016yjf,Yang:2018lew}. 
We hope to make further progress on these important goals in a near future.

The spin polarization in local equilibrium of conventional hydrodynamics description is fixed by hydrodynamic variables, such as temperature and vorticity, and it is not an independent hydrodynamic variable. However, since the total angular momentum including spin has to be conserved, one may think of formulating a hydrodynamics description of possible interplay between spin and orbital angular momenta.
There has been recent development in this ``spin hydrodynamics" in relativistic regime \cite{Florkowski:2017ruc,Florkowski:2018fap,Hattori:2019lfp}. Since spin is of order $\hbar$, one can think of this as a ${\cal O}(\hbar)$ quantum correction to the conventional classical hydrodynamics description. 
Some of the transport coefficients in this spin hydrodynamics \cite{Hattori:2019lfp} should in principle 
be determined by the quantum kinetic theory that we aim to construct.

\vskip 1cm \centerline{\large \bf Acknowledgment} \vskip 0.5cm

We thank Yukinao Akamatsu, Gokce Basar, Misha Stephanov and Derek Teaney for helpful discussions.
This work is supported by the U.S. Department of Energy, Office of Science, Office of Nuclear Physics, with the grant No. DE-SC0018209 and within the framework of the Beam Energy Scan Theory (BEST) Topical Collaboration.

\section*{Appendix: A simplification by rotational symmetry}

In this appendix, we show that the following integral is non-zero only when $s=s'$, and the value doesn't depend on $s$ either,
\be
\int {d^3\bm p'\over (2\pi)^3}{1\over 2 E_{p'}}\sum_{s''}[\bar u(\bm p,s)\gamma^\mu u(\bm p',s'')][\bar u(\bm p',s'')\gamma^\nu u(\bm p,s')] G_{\mu\nu}(\bm p-\bm p',t_a)e^{i(E_p-E_{p'})t_a}\,,\label{SEint}
\ee
when the gluon two-point function, $G_{\mu\nu}$, has a decomposition to longitudinal and transverse parts in Coulomb gauge,
\be
G_{\mu\nu}(t,\bm q)=\delta_{\mu 0}\delta_{\nu 0} G_L(t,q)+\Pi^T_{\mu\nu}(\bm q) G_T(t, q)\,,\quad q\equiv |\bm q|\,,
\ee
where the only non-zero elements of $\Pi^T_{\mu\nu}(\bm q)$ is
\be
\Pi^T_{ij}(\bm q)=(\delta_{ij}-\bm q^i\bm q^j/q^2)\,,\quad i,j=1,2,3\,.
\ee

Let us first show this for the longitudinal case. From the explicit expressions for $\gamma^\mu$ in (\ref{gammaM}), and the spinor in (\ref{spinor}) that we reproduce here
\be
 u(\bm p,s)=\left(\begin{array}{c} \sqrt{E_p-s p}\,\xi_s(\bm p) \\ \sqrt{E_p+sp}\,\xi_s(\bm p)\end{array}\right)\,,\label{spinornew}
\ee
we have (recall $\bar u=- u^\dagger \gamma^0$)
\bear
&&\sum_{s''} u(\bm p',s'') \bar u(\bm p',s'') \gamma^0=\sum_{s''}u(\bm p',s'')  u^\dagger(\bm p',s'') \nonumber\\
&=&\sum_{s''}\left(\begin{array}{cc} (E_{p'}-s'' p')\xi_{s''}(\bm p')\xi^\dagger_{s''}(\bm p') & m \xi_{s''}(\bm p')\xi^\dagger_{s''}(\bm p')\\ m\xi_{s''}(\bm p')\xi^\dagger_{s''}(\bm p') & (E_{p'}+s'' p')\xi_{s''}(\bm p')\xi^\dagger_{s''}(\bm p')\end{array}\right)\nonumber\\
&=&\left(\begin{array}{cc} E_{p'}-\bm p'\cdot\bm\sigma & m \\ m & E_{p'}+\bm p'\cdot\bm\sigma\end{array}\right)\,,\label{sumapp2}
\eear
where we use 
\be
\xi_{s''}(\bm p')\xi^\dagger_{s''}(\bm p')={\cal P}_{s''}(\bm p')={1\over 2}\left(1+s''{\bm p'\cdot\bm\sigma\over p'}\right)\,.
\ee
Now, consider the $\bm p'$ integral in (\ref{SEint}), and introduce spherical coordinates
$(p',\theta',\phi')$, where $\theta'$ is the azimuthal angle between $\bm p'$ and $\bm p$, and $\phi'$ is the polar angle around the perpendicular plane to $\bm p$.
It is easy to see, due to rotational symmetry of other parts in the integrand around $\phi'$, that only $\phi'$ dependence 
appears in the spinor sum (\ref{sumapp2}), as transverse components of $\bm p'$ with respect to $\bm p$. Since these transverse part of $\bm p'$ will integrate to zero after $\phi'$ integration,
it is clear that $\bm p'\cdot\bm\sigma$ in (\ref{sumapp2}) will be replaced by something proportional to $\bm p\cdot\bm\sigma$ after $\phi'$ integration. Therefore, (\ref{SEint}) becomes after $\bm p'$ integration
\be
u^\dagger(\bm p,s) \left(\begin{array}{cc} A+B(\bm p\cdot\bm\sigma) & C \\ C & A-B(\bm p\cdot\bm\sigma)\end{array}\right) u(\bm p,s')\,,\label{SEint2}
\ee
with some constants $A,B,C$. As $u(\bm p,s)\propto \xi_s(\bm p)$ and $(\bm p\cdot\bm\sigma)\xi_s(\bm p)=sp \xi_s(\bm p)$, and $\xi_s^\dagger(\bm p)\xi_{s'}(\bm p)=\delta_{s, s'}$,
we conclude that (\ref{SEint}), that is (\ref{SEint2}), is non-zero only when $s=s'$.
To show that the value doesn't depend on $s=s'$,
we use the explicit spinor (\ref{spinornew}) to evaluate (\ref{SEint2}) to obtain
\be
(E_p-sp)(A+spB)+(E_p+sp)(A-spB)+2m C=2(E_p A-p^2B+m C)\,,
\ee
which is indeed independent of the choice of $s$.

The proof in the transverse case is more complicated, but the idea is the same.
\bear
&&-\sum_{s''} \gamma^0\gamma^i u(\bm p',s'') \bar u(\bm p',s'') \gamma^j \Pi^T_{ij}(\bm q)\nonumber\\ &=&\sum_{s''}\left(\begin{array}{cc} \sigma^i & 0\\
0 & -\sigma^i\end{array}\right)u(\bm p',s'')  u^\dagger(\bm p',s'')\left(\begin{array}{cc} \sigma^j & 0\\
0 & -\sigma^j\end{array}\right) \left(\delta_{ij}-\hat{\bm q}^i\hat{\bm q}^j\right)\nonumber\\
&=&\left(\begin{array}{cc} \sigma^i & 0\\
0 & -\sigma^i\end{array}\right)\left(\begin{array}{cc} E_{p'}-\bm p'\cdot\bm\sigma & m \\ m & E_{p'}+\bm p'\cdot\bm\sigma\end{array}\right)\left(\begin{array}{cc} \sigma^j & 0\\
0 & -\sigma^j\end{array}\right)\left(\delta_{ij}-\hat{\bm q}^i\hat{\bm q}^j\right)\nonumber\\
&=&\left(\begin{array}{cc}  2E_{p'}+2\bm p'\cdot\bm\sigma-2i (\hat{\bm q}\cdot\bm\sigma)((\hat{\bm q}\times\bm p')\cdot\bm\sigma) & -2m\\ -2 m &  2E_{p'}-2\bm p'\cdot\bm\sigma+2i (\hat{\bm q}\cdot\bm\sigma)((\hat{\bm q}\times\bm p')\cdot\bm\sigma) \end{array}\right)\,.\nonumber\\\label{sumapp3}
\eear
Upon $\bm p'$ integration, we again have $\bm p'\cdot\bm\sigma$ become proportional to $\bm p\cdot\bm\sigma$, and recalling that $\bm q=\bm p-\bm p'$, we have
\be
(\hat{\bm q}\cdot\bm\sigma)((\hat{\bm q}\times\bm p')\cdot\bm\sigma)=
(\hat{\bm q}\cdot\bm\sigma)((\hat{\bm q}\times\bm p)\cdot\bm\sigma)=\epsilon^{jkl}\hat{\bm q}^i\hat{\bm q}^k \bm p^l \sigma^i\sigma^j\,.
\ee
Due to symmetry of $\phi'$ and $\bm q=\bm p-\bm p'$, the $\hat{\bm q}^i\hat{\bm q}^k$ part will become, after $\phi'$ integration, a linear combination of $\delta^{ik}$ and $\bm p^i\bm p^k$. 
Obviously, these two structures are the only possible rank-2 structures with $\bm p$, since the only available vector after $\bm p'$ integration is $\bm p$. The $\bm p^i\bm p^k$ piece doesn't contribute to the above due to $\epsilon^{jkl}\bm p^l$, while the $\delta^{ik}$ piece results in 
\be
\epsilon^{ijl}\sigma^i\sigma^j\bm p^l\sim \bm p\cdot\bm\sigma\,,
\ee
that is, the same $\bm p\cdot\bm\sigma$ structure.
Therefore, (\ref{sumapp3}) becomes, after $\bm p'$ integration in (\ref{SEint}),
\be
 \left(\begin{array}{cc} A'+B'(\bm p\cdot\bm\sigma) & C' \\ C' & A'-B'(\bm p\cdot\bm\sigma)\end{array}\right)\,,
 \ee
 with constants $A',B',C'$, that is the same structure we obtain in the longitudinal polarization case (see (\ref{SEint2})), and hence the same conclusions follow.


 \vfil

\end{document}